\documentstyle[12pt,aaspp4]{article}

\def\linebreak{\hfil\break}

\def\
{\hfil\linebreak}

\def\etal{{\it et al}. }
\def\degree{\ifmmode {^\circ}\else {$^\circ$}\fi}
\def\mum{\ifmmode {\rm \mu {\rm m}}\else $\rm \mu {\rm m}$\fi}
\def\arcsec{\ifmmode ^{\prime \prime}\else $^{\prime \prime}$\fi}

\def\inch{\ifmmode ^{\prime \prime}\else $^{\prime \prime}$\fi}
\def\arcmin{\ifmmode ^{\prime}\else $^{\prime}$\fi}

\def\rsun{\ifmmode {\rm R_{\odot}}\else $\rm R_{\odot}$\fi}
\def\lsun{\ifmmode {\rm L_{\odot}}\else $\rm L_{\odot}$\fi}
\def\msun{\ifmmode {\rm M_{\odot}}\else $\rm M_{\odot}$\fi}
\def\msunyr{\ifmmode {\rm M_{\odot}~yr^{-1}}\else $\rm M_{\odot}~yr^{-1}$\fi}
\def\mdot{\ifmmode {\rm \dot{M}}\else $\rm \dot{M}$\fi}
\newbox\grsign \setbox\grsign=\hbox{$>$} \newdimen\grdimen \grdimen=\ht\grsign
\newbox\simlessbox \newbox\simgreatbox
\setbox\simgreatbox=\hbox{\raise.5ex\hbox{$>$}\llap
     {\lower.5ex\hbox{$\sim$}}}\ht1=\grdimen\dp1=0pt
\setbox\simlessbox=\hbox{\raise.5ex\hbox{$<$}\llap
     {\lower.5ex\hbox{$\sim$}}}\ht2=\grdimen\dp2=0pt

\def\simless{\mathrel{\copy\simlessbox}}
\begin{document}

\centerline{\Large {\bf Flickering in FU Orionis}}
\vskip 7ex
\centerline{Scott J. Kenyon}
\centerline{Harvard-Smithsonian Center for Astrophysics}
\centerline{60 Garden Street, Cambridge, MA 02138 USA} 
\centerline{e-mail: skenyon@cfa.harvard.edu}
\vskip 2ex
\centerline{E. A. Kolotilov}
\centerline{Crimean Laboratory}
\centerline{Sternberg State Astronomical Institute, p/o Nauchny, 334413 Crimea}
\centerline{e-mail:  kolotilov@sai.crimea.ua}
\vskip 2ex
\centerline{M. A. Ibragimov}
\centerline{Ulugh Beg Astronomical Institute, Academy of Sciences of Uzbekistan}
\centerline{Astronomicheskaya ul. 33, Tashkent, 700052 Uzbekistan}
\centerline{e-mail:  mansur@astrin.uzsci.net}
\vskip 1ex
\centerline{and}
\vskip 1ex
\centerline{Janet A. Mattei}
\centerline{American Association of Variable Star Observers}
\centerline{25 Birch Street, Cambridge, MA 02138 USA}
\centerline{e-mail:  jmattei@aavso.org}
\received{6 August 1999}
\accepted{21 October 1999}
%


\begin{abstract}

We analyze new and published optical photometric data of FU Orionis,
an eruptive pre--main-sequence star.  The outburst consists of
a 5.5 mag rise at B with an $e$-folding timescale of $\sim$ 50 days.
The rates of decline at B and V are identical, 0.015$\pm$0.001 mag yr$^{-1}$.
Random fluctuations superimposed on this decline have an amplitude 
of 0.035 $\pm$ 0.005 mag at V and occur on timescales of 1 day or 
less.  Correlations between V and the color indices U--B, B--V, and 
V--R indicate that the variable source has the optical colors of a 
G0 supergiant.  We associate this behavior with small amplitude 
flickering of the inner accretion disk.

\end{abstract}

\subjectheadings{accretion disks --
stars pre--main-sequence --
stars: formation --
stars: individual (FU Orionis) }

\section{INTRODUCTION}

FU Orionis objects -- sometimes known as FUors -- are eruptive 
pre--main-sequence stars located in active star forming regions
(\cite{her66}, 1977; \cite{har96}; \cite{ken99}).  Roughly half of the 
11 commonly accepted FUors have been observed to rise 3--5 mag 
in optical or near-IR brightness on timescales of 1--10 yr.  
Other FUors have been identified based on properties similar
to eruptive FUors, including 
(i) absorption features of F--G supergiants on optical 
spectra and K--M giants on near-IR spectra (\cite{her77}; 
\cite{mou78}; \cite{car87}; \cite{sto88}; \cite{sta92}); 
(ii) large excesses of radiation over normal F--G stars at 
ultraviolet, infrared, submillimeter, and centimeter wavelengths 
(\cite{wei89}, 1991; \cite{ken91}; \cite{rod90}; \cite{rod92});
(iii) distinctive reflection nebulae (\cite{goo87}); and
(iv) clear association with optical jets, HH objects, and 
molecular outflows (\cite{rei91}; \cite{eva94}).

FUor eruptions are often accepted as accretion events in the disk 
surrounding a low-mass pre--main-sequence star (\cite{har85}; 
\cite{lin85}; \cite{har96}; for alternative interpretations,
see Herbig \& Petrov 1992; Petrov \etal 1998). In this picture, 
the accretion rate 
through the disk increases by 2--3 orders of magnitude to 
$\sim 10^{-4}~\msunyr$.  In addition to providing energy for
the luminosity increases of FUors, this model naturally explains
the broad spectral energy distributions, the variation of rotational
velocity and spectral type with wavelength, and color changes 
during the optical decline of V1057 Cyg, among other observed 
properties.  

Despite the success of the disk hypothesis, one observable 
characteristic of disk accreting systems -- flickering -- has 
not been observed in any known FUor.  In most systems with luminous
accretion disks, flickering is observed as a series of random 
brightness fluctuations with amplitudes of 0.01--1.0 mag that
recur on dynamical timescales (\cite{rob76}; \cite{bru92}).  
Often accepted as a `signature' of disk accretion, flickering 
is believed to be a dynamical variation of the energy output 
from the disk\footnote{In the cataclysmic variables (short 
period binary systems with an accretion disk surrounding a white 
dwarf), random flickering occurs on timescales of seconds to 
minutes.  Dwarf nova oscillations and quasi-periodic oscillations 
are semi-coherent periodic variations observed on similar timescales. 
Flickering is also occasionally associated with material in the 
`bright spot' at the outer edge of the disk (see \cite{war95}).
It is unclear whether or not these classes have distinct analogs 
among accreting pre-main sequence stars. In this paper, we 
use flickering to distinguish rapid variations of light from 
the inner disk from variations of the bright spot.}.
Thus, it provides some measure of the 
fluctuations in the physical structure of the disk and might
someday serve as a diagnostic of physical properties within 
the disk (\cite{bru93}; \cite{bru94}; \cite{war95}).

In this paper, we search for evidence of flickering in the historical 
light curve of FU Ori.  In addition to the outburst, we find good
evidence for small-amplitude brightness fluctuations on a timescale
of 1 day or less.  Color changes correlated with the brightness changes
indicate a variable source with the optical colors of a G0 supergiant.
The amplitude, color temperature, and timescale of the variations have 
much in common with the flickering observed in short period interacting
binary systems.  After
ruling out several possible alternatives, we conclude that flickering
is the most likely interpretation for short-term variability in FU Ori.
The most plausible location for the flickering source is the inner
edge of the disk, where temperatures lie between the stellar temperature
and the maximum disk temperature of $\sim$ 7000 K.

We describe the observations in \S2, analyze the light curve in \S3,
and conclude with a brief discussion and summary in \S4.

\section{OBSERVATIONS}

We acquired UBV photometry of FU Ori with the 60-cm Zeiss reflector
at the Crimean Laboratory of the Sternberg State Astronomical Institute.
The observations usually were made through a 13\arcsec~aperture; 
a 27\arcsec~aperture was used on nights of poor seeing.  These data
were reduced using BD+8\degree1051 as the comparison star and other
nearby stars as controls (see \cite{kol85}).  Table 1 lists the 
results. The uncertainty in the calibration is $\pm$0.01--0.02 mag 
for V and B--V, and $\pm$0.02--0.04 mag for U--B.  

We supplement the UBV data with additional photoelectric photometry 
from the Maidanak High Altitude Observatory.  Ibragimov (1997) 
describes UBVR data acquired during 1981--1994 as part of the 
ROTOR project.  The data have been reduced 
to the standard UBVR system with typical errors of $\pm$0.015 mag 
for V and V--R, $\pm$0.02 mag for B--V, and $\pm$0.04--0.08 mag 
for U--B.

We also consider visual observations of FU Ori compiled by the
American Association of Variable Star Observers.  The error of
a typical estimate is 0.1--0.2 mag using standard stars in the field
of FU Ori calibrated from photoelectric observations.  We simplify 
a comparison with photoelectric data by computing twenty day means of 
the over 7000 AAVSO observations.  With $\sim$ 15 observations
per twenty day interval, this procedure reduces the typical error of
a twenty day mean to $\sim$ 0.03 mag, only slightly larger than the
quoted error of the photoelectric data.

Finally, we obtained low resolution optical spectra of FU Ori during 
1995--1998 with FAST, a high throughput, slit spectrograph mounted 
at the Fred L. Whipple Observatory 1.5-m telescope on Mount Hopkins, 
Arizona (\cite{fab98}).  We used a 300 g mm$^{-1}$ grating  blazed 
at 4750 \AA, a 3\arcsec~slit, and recorded the spectra on a thinned
Loral 512 $\times$ 2688 CCD.  These spectra cover 3800--7500 \AA~at 
a resolution of $\sim$ 6 \AA.  On photometric nights, we acquired 
standard star observations to reduce the FU Ori data to the 
Hayes \& Latham (1975) flux scale using NOAO IRAF\footnote{IRAF 
is distributed by the National Optical Astronomy Observatories,
which is operated by the Association of Universities for Research in 
Astronomy, Inc., under contract to the National Science Foundation}. 
These calibrations have an accuracy of $\pm$0.05 mag.  This uncertainty
is comparable to the probable error in spectrophotometric data 
acquired with the Kitt Peak National Observatory Intensified
Reticon Spectrograph reported in Kenyon \etal (1988).  

Table 2 lists indices for several absorption lines using O'Connell's
(1973) definition, $I_{\lambda} = F_{\lambda}/\bar{F}_{\lambda}$,
where $F_{\lambda}$ is the measured flux in a 20 \AA~bandpass 
centered at wavelength $\lambda$ and $\bar{F}_{\lambda}$ is the 
continuum flux interpolated from continuum bandpasses on either 
side of the absorption feature.  Repeat measurements indicate an
error of $\pm$ 0.03 mag for each index.

\section{LIGHT CURVE ANALYSIS}

The lower panel of
Figure 1 shows historical B+photographic and optical light curves 
for FU Ori.  Following a 5--6 mag rise at B, the system has 
declined by $\sim$ 1 mag in nearly 70 yr.  The decline in visual 
light has closely followed the B light curve for the past 30 yr.
In addition to a long-term wave-like variation, the brightness 
shows considerable scatter of roughly 0.1 mag at BVR and almost 
0.2 mag in U on timescales of days to months.  These fluctuations
are much larger than the quoted photometric errors.  Ibragimov (1997) 
commented on fluctuations in brightness {\it and} color indices 
throughout 1981--1994.  Similar variations are visible in light 
curves of V1057 Cyg and V1515 Cyg (\cite{ibr97}).  We plan analyses
of these other FUors in future publications.

\subsection{Outburst Model}

To analyze brightness fluctuations in FU Ori, we begin by 
separating long-term changes with timescales of years from 
shorter timescale variations.  We consider a simple model 
for the outburst

\begin{equation}
m = m_0 + m_{rise}(t - t_0)
\end{equation}

\noindent
where $m_0$ is the average quiescent magnitude and
$t_0$ is the time of the outburst.
The brightness during outburst is a function of time, $t$,

\begin{equation}
m_{rise} = \left\{ \begin{array}{l c r}
           0 & &t < t_0 \\
           \delta m_{rise} \{1 - e^{-(t - t_0) / \tau_{rise}} \} + \dot{m} (t - t_0) & \hspace{5mm} & t \ge t_0 \\
           \end{array}
           \right .
\end{equation}

\noindent
where $\delta m_{rise}$ is the amplitude of the outburst,
$\tau_{rise}$ is the e-folding time of the rise to maximum, and
$\dot{m}$ is the rate of decline from maximum.
We derive parameters for this model using a downhill simplex 
method to minimize the function

\begin{equation}
\chi^2 = \sum_{i=1}^N [(m - m_i)/\sigma_i]^2 ~ ,
\end{equation}

\noindent
for $N$ observations $m_i$ having uncertainty $\sigma_i$.  
We derive the model parameters
$m_0$, $\delta m_{rise}$, $\dot{m}$, and $\tau_{rise}$ for 
an adopted $t_0$ and vary $t_0$ separately to produce a
minimum in $\chi^2$.  This procedure works better than fitting
all parameters at once due to the sparse nature of the light
curve for $t \lesssim t_0$.  To estimate errors of the model
parameters, we compute residuals $r_i$ of the light curve about 
the fit, construct new light curves by adding gaussian noise 
with amplitude $r_i$ to the data, and extracting new model
parameters.  We adopt as best the median values for model
parameters from 10,000 such trials; the quoted errors are 
the inter-quartile ranges of each model parameter.

Table 3 summarizes results of model fits to the light curves
in Figure 1.  The sparse data before outburst yield a poor 
measure of the pre-outburst brightness and the start of the 
outburst.  We estimate $t_0$ = JD 2428497 $\pm$ 40 and 
$m_0 = 15.55 \pm 0.40$ from the B-band data; these errors set 
the uncertainties in the other model parameters.  The uncertainty 
in the decline rate at B is small;  the visual data yield a nearly 
identical rate of decline despite lack of data near maximum.  

The spectroscopic data show little evidence for any change 
in the mean optical spectral type during the recent 0.3
mag decline in the mean V brightness.
The Mg~I index, which tracks the spectral type rather well,
increased by at most 0.02 mag during these years.  The H$\beta$
and Na~I indices, which should measure the wind of FU Ori,
have also remained constant.  We show below that the mean
colors, U--B, B--V and V--R, of FU Ori have changed by 
$\lesssim$ 0.02 mag in 15 years.  The constant colors and 
spectral indices indicate that the optical spectral type has
remained constant to within $\pm$ 1 subclass since 1985.

The upper panel of Figure 1
shows residual light curves about the best-fit model.
The rms dispersion about the light curves is
0.06 mag for photoelectric data,
0.11 mag for visual data, and
0.24 mag for photographic data.
All are large compared to typical uncertainties.

Some of the scatter in the residual light curves is due to a 
long-term wave with an amplitude of several tenths of a magnitude. 
The residual B light curve has three prominent crests at $\sim$ 
JD 2428500, JD 2432000, and JD 2436000, with a possible small
crest at $\sim$ JD 2433500.  This behavior appears to vanish in
the B light curve at later times.  However, the residual visual 
light curve then shows a similar wave-like oscillation with crests 
at $\sim$ JD 2439000 and JD 2442500.  Another crest at $\sim$
JD 2447000 is present in the photoelectric V light curve shown 
in Figure 2 and described below. These peaks are roughly in
phase with the residual B light curve for periods of $\sim$ 4000
days.  

There are considerable photometric variations in addition to the 
wave.  These occur on short timescales, tens of days or less, and 
have amplitudes that seem uncorrelated with the overall system 
brightness.  We will 
analyze these in the next section and then consider possible
origins for the short-term and long-term variations.

Despite the large residuals, the model fits leave negligible 
trends in the data.  The median Spearman rank correlation 
coefficient between the V brightness and time for the 10,000 
Monte Carlo trials is 0.87 for the B + photographic data and
0.84 for the visual data.  The median slope of a linear fit 
to the residual light curves is $\simless 10^{-4}$ mag yr$^{-1}$ 
for the B + photographic data and $\simless 10^{-6}$ mag yr$^{-1}$ 
for the visual data.  We conclude that the simple model provides
a reasonable fit to the long-term light curve and now consider 
the nature of the shorter timescale fluctuations.

\subsection{Fluctuation Timescale}

To test whether or not the short timescale photometric variations
in FU Ori are flickering, we need to verify that 
(i) they are real changes in the brightness and color of FU Ori,
(ii) they occur on short timescales but are not periodic, and
(iii) they can be plausibly associated with radiation from the disk.
We establish the reality of the variations in two parts.
We first demonstrate that the variations occur on short timescales
with amplitudes larger than the photometric errors.  We show 
later that color changes correlate with brightness changes.

We consider a Monte Carlo model for the photoelectric data 
described in \S2.  The lower panel of Figure 2 shows V-band data
acquired during the last 15 yr.  Both the wave-like oscillation
and the large scatter about this oscillation are apparent.  The
solid line in this Figure is the seasonal mean light curve,
the average brightness for each year of observation.  The top 
panel in the Figure is the difference between the actual data 
and the seasonal mean.  This residual light curve has a large 
amplitude but no linear trend or wave-like feature.

Our analysis indicates a small periodic component in residual 
light curves for B and V data.  Periodograms suggest a period 
of 17 $\pm$ 1 days in the B data, which has been reported 
previously (\cite{kol85}).  This periodic component 
has an amplitude of 0.009 $\pm$ 0.003 mag.  The V data has a
best period of 111.4 $\pm$ 1.6 days with an amplitude of
0.012 $\pm$ 0.003 mag.  There is no indication of the 17 day
period in the V light curve.  The amplitudes of these `periodicities'
are comparable to the photometric errors but small compared to
the amplitude of the fluctuations in the residual light curves
(see Figures 1--2). 

To measure the amplitude of the non-periodic component in the
residual light curve, we use a Monte Carlo model.  We replace
each observation V$_i$ with a random brightness having amplitude
$a_v$ and offset $v_0$,

\begin{equation}
v_i = v_0 + a_v g_i ~ ,
\end{equation}

\noindent
where $g_i$ is a normally distributed deviate with zero mean
and unit variance (\cite{pre92}).  Artificial light curves that
provide a good match to the actual light curve should have the
same amplitude and offset.  We quantify a `good match' by 
comparing the magnitude distributions of actual and artificial
light curves using the Kolmogorov-Smirnov (K-S) test.  We
reject poor matches with a low probability of being drawn 
randomly from the same distribution as the data.  The `best' 
match maximizes the median K-S probability from 10,000 trials.  
We establish error estimates for the best parameters and the 
proper scale for this measure by comparing two artificial light 
curves generated in each of 10,000 trials.

This procedure yields best parameters of 
$v_0 = 0.0016 \pm 0.0009$ mag and
$a_v = 0.033 \pm 0.005$ mag for the residual V light curve.
Artificial light curves with these parameters have a high
probability, 68\% or larger, of being drawn randomly from
the same distribution as the actual data.  The offset of the 
model light curve is consistent with zero.  The amplitude of the
model is roughly twice the quoted 1$\sigma$ error of 0.015 mag.

The artificial data sets have periodic variations similar to
the real data sets.  Periods of 10--100 days are common in the 
artificial B and V light curves.  Mean light curves folded on these
periods have amplitudes, 0.01 $\pm$ 0.003 mag, similar to 
those quoted above for the real data.  The `best' period 
is different in each artificial data set, but the amplitude
is nearly constant.  This amplitude is small compared to the 
random component of the fluctuation.  We suspect the periodic 
variations are due to the sampling of the light curve.  Such
`periodicities' illustrate some of the dangers of period analysis. 

There are two explanations for the 0.03 mag variations in the
residual V light curve.  A simple test should distinguish real
fluctuations from measurement error.  Real fluctuations in V
should be accompanied by correlated variations in the color
indices, U--B, B--V, and V--R.  Color variations should be
uncorrelated with the V brightness and with each other if the
measurement error is 0.03 mag in V instead of 0.015 mag.
In either case, the Monte Carlo model demonstrates that the
fluctuations occur on short timescales.  The typical observation
frequency is $\sim$ 1 day$^{-1}$ (344 out of 663 observations),
0.5 day$^{-1}$ (82 observations), or 0.333 day$^{-1}$ (54
observations).  Fluctuations must occur on timescales 
$\lesssim$ 1 day based on the success of the Monte Carlo 
model in reproducing the light curve.

\subsection{Color Variations}

Figure 3 shows the variations in the optical colors as a function
of time.  The solid line in each of the left-hand panels is 
the seasonal mean color for the individual data points.  There 
is little evidence for a substantial color change in FU Ori 
during a 0.3 mag decline in V.
The seasonal means for B--V and V--R are constant to within the
photometric errors.  The variation in the mean U--B color is roughly
twice the photometric error but shows no obvious trend with time.

Despite the lack of long-term variation in the colors, there are
considerable short-term variations.  The right hand panels in 
Figure 4 show the color fluctuations about the seasonal means 
in the left hand panels.  The full amplitudes of the color 
variations are $\sim$ 0.2 mag in U--B, and $\sim$ 0.1 mag 
in B--V and V--R.  These variations are 3--5 times the 
quoted photometric errors.

Figures 4--5 show the correlation of color variations with the 
V-band fluctuations analyzed in \S3.2.  To construct these plots, 
we derived separate seasonal means for the V-band observations 
associated with each color observation (because a given color was
not always obtained with each V measurement) and subtracted the
appropriate seasonal mean from each data point.  The Spearman 
rank correlation coefficient is 
1.3 $\times~10^{-2}$ for 438 $\delta$(U--B)--$\delta$V pairs,
2.6 $\times~10^{-9}$ for 626 $\delta$(B--V)--$\delta$V pairs, and
2.9 $\times~10^{-11}$ for 622 $\delta$(V--R)--$\delta$V pairs.
We fit each correlation with a straight line using the Press \etal
(1992) subroutine FITEXY, assuming that the 1$\sigma$ errors in
each coordinate are equal to the quoted photometric errors. The
results yield:

\begin{equation}
\delta (U-B) = -0.0002 \pm 0.0034 - (0.40 \pm 0.14) \delta V
\end{equation}

\begin{equation}
\delta (B-V) = (-0.24 \pm 8.0) \times 10^{-4} - (0.12 \pm 0.02) \delta V
\end{equation}

\begin{equation}
\delta (V-R) = (-0.015 \pm 6.1) \times 10^{-4} + (0.15 \pm 0.02) \delta V
\end{equation}

\noindent
The correlation between $\delta$(U--B) and $\delta$V is weak;
U--B becomes redder as the source becomes brighter.  Both
$\delta$(B--V) and $\delta$(V--R) correlate well with $\delta$V;
as the source brightens, B--V becomes redder while V--R becomes
bluer.

The slopes of the short-term color variations in equations (5)--(7) 
differ from the apparent slope of the long-period wave in the B
and visual light curves.  The lack of a clear B variation associated
with the visual wave suggests $\delta (B-V) \approx C \delta V$,
with $C \approx$ 1.  This behavior suggests that the wave and the
short-term fluctuations have different physical origins.  Analysis
of photoelectric data with a longer time baseline is needed to
verify this point.

To test the accuracy of the correlation coefficients and measured
slopes, we constructed artificial color curves using the Monte Carlo
model described in \S3.2.  We matched model color curves having 
amplitudes similar to the observations to model V-band light curves, 
measured the correlation coefficients, and derived the slopes of
straight line fits to the artificial residuals.  We repeated this
exercise using known correlations of color index with brightness.
Random light curves yield no correlation between color index and
brightness.  Correlated changes in color index with brightness
yield the measured correlation coefficients if the amplitude of
the brightness variation is $a_v$ = 0.036 $\pm$ 0.007, if the slopes
and 1$\sigma$ errors of the color variations are those quoted in 
Equations (5)--(7), and if the photometric errors are
0.015 $\pm$ 0.01 in V, 0.02 $\pm$ 0.01 in B--V and V--R, 
and 0.07 $\pm$ 0.02 in U--B.

The small color variations make it difficult to find robust 
correlations among the color indices.  There is a 3$\sigma$
correlation between U--B and B--V, with a Spearman rank 
coefficient of $4.7~\times~10^{-3}$.  The Spearman rank
correlation coefficient is much larger, 0.25, for B--V and V--R. 
To test the importance of these results, we repeated these tests 
using artificial data sets with known correlations between the
color indices.  The correlation between $\delta$(U--B) and 
$\delta$(B--V) is always detected in data with small photometric 
errors; the Spearman rank correlation coefficient is $10^{-3}$
or smaller in all of our trials.  The correlation between 
$\delta$(B--V) and $\delta$(V--R) has 2$\sigma$ or smaller
significance in all of our trials. Reducing the photometric 
errors by factors of 2--3 allows us to recover known correlations
at the 3$\sigma$ level.

These results provide good evidence that the observed variations 
in brightness and color are intrinsic to FU Ori.  The correlations
between the B--V or V--R color and the V brightness are robust.
The measured correlation coefficients are reasonable given the
magnitude of the color changes and the photometric errors.
We next consider the physical nature of the brightness changes
and then compare these properties with the flickering observed
in other accreting systems.

\subsection{Physical Nature of Light and Color Variations}

The observed optical light and color variations in FU Ori
are small, $\sim$ 0.035 mag in V, $\sim$ 0.015 mag in U--B,
and $\sim$ 0.004 mag in B--V and V--R.  To understand the 
origin of this variability, we consider the observed colors
as small perturbations about the colors of the `average'
source in FU Ori.  We adopt the mean colors of the system 
as the average colors,

\begin{equation}
U-B = 0.84 ~~~~~~~~ B-V = 1.35 ~~~~~~~~ V-R = 1.15 ~~~~~.
\end{equation}

\noindent
and correct these colors for interstellar reddening assuming 
a standard extinction law (\cite{mat90}) and $A_V$ = 
2.2 mag (\cite{ken88}),

\begin{equation}
(U-B)_0 = 0.33 ~~~~~~~~ (B-V)_0= 0.64 ~~~~~~~~ (V-R)_0 = 0.60 ~~~~~.
\end{equation}

\noindent
Figure 6 compares the reddening-corrected color indices of FU Ori
with colors for F and G supergiants as indicated in the plot.
The average colors of FU Ori, shown as the box, are offset from
standard stellar loci.  The solid line in each panel shows how 
the colors change as the source brightens; the colors move away 
from the supergiant locus if the source fades in brightness.
Both lines indicate that the best stellar match to the changes
in brightness and color is a G0 supergiant.  The dashed lines 
indicate how the colors change if the slopes of the relation
between the V-band brightness and the color indices are 
$\pm 2 \sigma$ different from those derived in Equations (5)--(7).
This result shows that the uncertainty in the stellar match 
is less than one spectral subclass.

The best stellar match to the color variations is correlated
with the adopted reddening.  An F5 supergiant can match the
observed color variations for $A_V$ = 3.2 mag; a G5 supergiant
is the best match for $A_V$ = 1.2 mag.  Intrinsic colors for
$A_V \lesssim$ 1.5 mag and $A_V \gtrsim$ 3 mag are inconsistent 
with the G-type optical spectrum.  This constraint limits the
spectral type of the variable source to F7--G3.

To find a reasonable explanation for the color variations,
we first consider mechanisms appropriate for single stars
without circumstellar disks.  Although we believe an accretion
disk provides the best model for observations of all FUors,
it is important to consider alternatives before we identify
flickering as the source of the variability in FU Ori.  
Obscuration, pulsation, and rotation are obvious choices for 
short period variations with small amplitudes in an isolated 
star. Random obscuration events by small, intervening dust 
clouds near the central source are a popular model for rapid
variations in Herbig Ae/Be stars and other pre-main sequence
stars (e.g., \cite{nat97}; \cite{ros97}; and references therein).  
The observed pulsational amplitude of the pre-main sequence 
star HR 5999, 0.013 mag (\cite{kur95}), is close to the 
amplitude observed in FU Ori. The period of 5 hr is short 
compared to the spacing of our data; a similar period 
in FU Ori might well be missed by the analysis described 
above.   Rotational modulations with small amplitudes and 
periods of 1--2 days are observed in many pre--main-sequence 
stars (\cite{kea97} and references therein) and could 
plausibly produce a similar variation in FU Ori.  

Despite its attractiveness, a stellar pulsation in FU Ori seems 
unlikely.  With an effective temperature of $\sim$ 6000--6500 K 
and a luminosity of $\sim$ 200 \lsun~(e.g., \cite{ken99}), FU Ori 
lies within the classical instability strip (see \cite{gau95}). 
The bright Cepheid $\alpha$ UMi (Polaris) has a comparable amplitude 
and other Cepheids have similar periods.  Recent period-luminosity 
relations for Cepheids (\cite{fea97}; \cite{hin89}) yield periods
of $\sim$ 0.8 days for FU Ori if d = 500 pc and $A_V$ = 2.2 mag.
Although this period is reasonably consistent with our observations,
FU Ori lies well above the {\it pre--main-sequence} instability strip 
in the HR diagram (\cite{mar98}).  To test the possibility that a
pulsational instability might occur anyway, we constructed artificial
light curves with amplitudes of 0.035 mag and periods of 0.3--2.0 days.
We sampled these light curves in time as in our real observations 
and added noise equivalent to our photometric uncertainties.  Our
failure rate for recovering known periods in the artificial light 
curves is small, $\simless 10^{-3}$, for all periods considered,
based on 10,000 artificial light curves at each of 20 periods 
between 0.3 and 2.0 days.  The failure rate is largest for periods 
of 0.5, 1.0, and 1.5 days due to the $\sim$ 1 day spacing 
of the light curve.  
The failure rate decreases to $10^{-4}$ or less at other periods.
Our inability to detect any periodicity in the FU Ori light curve
and the lack of a theoretical instability strip for the observed
luminosity and temperature of FU Ori is a strong indication that 
pulsations do not produce the observed variation.

Stellar rotation is also an unlikely source of the variability.  
The light curve analysis for short periodicities rules out a 
rotational modulation of the light curve, unless the dark or
bright spots responsible for the variations vary in size or
intensity on timescales of several rotational periods.  At the
1--2 day periods that seem most plausible, FU Ori currently 
rotates close to breakup. If the star conserved angular momentum 
during the rise to maximum, the pre-outburst rotational velocity 
would have exceeded the breakup velocity by a factor of $\sim$ 3 
(see also \cite{har85}). 

Obscuration events similar to those envisioned in Herbig Ae/Be
stars (see \cite{ros97} and references therein) require special 
circumstances to explain the variability. 
Reddening by small dust clouds requires unusual particles to
account for a very steep reddening law, $R_V \sim$ 8, that
changes sign at V.  Light scattered off a reflection nebula 
might account for the different color variations of B--V and
V--R, but it is difficult to derive as steep a color variation 
as is observed with simple geometries and a wide range of dust 
properties.  Photopolarimetry would test this conclusion.  

Finally, fluctuations in the wind of FU Ori are a less plausible 
source of brightness changes than flickering.  Errico \etal (1997)
have suggested that the continuum optical depth through the
wind exceeds unity in the Balmer and Paschen continua. Small
variations in the optical depth due to inhomogeneities in the
outflow might account for small brightness changes.  If correct,
this hypothesis would predict a decrease in the amplitude of the 
variation with decreasing wavelength, because the optical depth
in the Paschen continuum decreases with decreasing wavelength.  
For reasonable
wind temperatures and densities, $\sim$ 5000--10000 K and $\sim$
$10^{10}$--$10^{14} \rm ~ cm^{-3}$ (\cite{cal93}; \cite{har95}),
the increase in the optical depth from B to R is roughly a factor 
of 3 for a gas in LTE.  There is no evidence for this behavior
in FU Ori.

We conclude that fluctuations associated with a stellar photosphere
or wind from the disk provide a poor explanation for the rapid, 
small-amplitude photometric variations in FU Ori.  The variations 
in FU Ori have 
much in common, however, with the flickering observed in other 
accreting systems, such as cataclysmic variables (CVs) and low 
mass X-ray binary systems (\cite{bru92}, 1994; \cite{bru93}; 
\cite{war95} and references 
therein).  In most CVs, 0.01--1 mag fluctuations occur on the 
dynamical timescale, seconds to minutes, of the inner disk.  Within 
each flicker, a CV becomes bluer as it gets brighter.  The large 
color temperature, $\gtrsim$ 20,000 K, of the variable source also 
plausibly confines the flickering to the inner disk in many CVs.  
Despite the lack of a good physical model for flickering in CVs, 
it clearly probes physical conditions in the inner disk.

The observed variations of FU Ori are also plausibly associated 
with the inner regions of a circumstellar accretion disk.  The 
timescale of the variation, $\lesssim$ 1 day, is close to the 
dynamical timescale of the inner disk, $\sim$ 0.1 day for a 
1 \msun~central star with a radius of 4 \rsun.  The temperature 
of the variable source, $\sim$ 6000 K for a G0 supergiant, is 
comparable to the inner disk temperature of 6500 K derived from
detailed fits to the spectral energy distribution and the profiles
of various absorption lines (\cite{ken88}; \cite{bel95};
\cite{tur97}).  Finally, 
the amplitude of the variation is similar to that observed in other 
accreting systems.  Given this behavior, we believe that the 
variations in FU Ori {\it are} flickering and thus provide 
additional evidence for an accretion disk in this system.

The main alternative to a variable accretion disk in FU Ori 
is variations associated with a magnetic accretion column.
Despite the success of this model in other pre--main-sequence
stars (e.g., \cite{gul96}), truncating an accretion disk with
the large accretion rate, $\sim 10^{-4} \msunyr$, estimated in 
FU Ori requires a large magnetic field, $\sim$ 10 kG. Limits
on the magnetic fields of other pre--main-sequence stars are 
much smaller, $\lesssim$ 2--3 kG (\cite{joh99}).  A much larger
field in FU Ori is unlikely.  The temperature of the variable 
source in FU Ori, $\sim$ 6000 K, also seems too cool to be 
associated with a magnetic accretion column, where the typical 
temperature is $\sim 10^4$ K (e.g., \cite{lam98}; \cite{cal99}).  
Both of these arguments make a stronger case for flickering as 
the source of the variability in FU Ori.

To see what we can learn about the inner disk from the variations,
we consider several simple models for the flickering.  We adopt 
the observed colors -- U--B, B--V, and V--R -- as the colors of
the average state of the disk.  We assume several sources of
variability in a steady-state disk,
(i) discrete changes in the mass accretion rate, $\mdot$, 
through the entire disk,
(ii) random fluctuations in the flux from any annulus in the
disk, and
(iii) random fluctuations in the flux from specific annuli
in the disk.
We chose a steady-state disk as the average source, because
steady disks provide a reasonable fit to the complete spectral 
energy distribution of FU Ori. Several experiments with non-steady
disks yield similar results.

Models where the entire disk can vary in brightness do not 
reproduce the observations.  We assume a disk composed of discrete
annuli with width $\delta R$ at a distance $R$ from the central star,
with $\delta R \ll R$.  Each annulus radiates as a star with the
effective temperature assigned to the annulus. The arrows in 
Figure 6 indicate the color 
variations produced by models where the flux from each annulus 
is a random fraction, between 1.0 and 1.1, of the flux from the 
annulus of a steady-state disk.  The color variation of the model 
clearly fails to account for the observed color variation.  
Allowing all annuli to vary coherently also produces color 
variations that disagree with the observed variation.

Successful models allow only specific parts of the disk to vary
in brightness. If disk annuli with the colors of F9--G1 supergiants
are the only annuli that vary, the color variation of the model 
follows the slope of the observed variation.  

Our results for flickering in FU Ori are at odds with predictions
of the simple steady-state accretion disk model.  In steady disk 
models for FUors, the disk temperature rises rapidly from zero at 
the stellar photosphere, $R = R_{\star}$, to $T_{max} \approx$ 
6500--7000 K at $R = 1.36~R_{\star}$ and then decreases radially 
outward (\cite{ken88}).  The G0 temperature of the flickering 
source has a temperature of $\sim$ 6000 K.  Disk material with 
this temperature has $ R = 2.5~R_{\star}$ in the steady model.  
Fluctuations in the energy output of this region, and the lack 
of fluctuations in hotter disk material, seem unlikely.  If we 
associate flickering with small changes in the mass accretion 
rate through the disk, $\mdot$, or in the scale height of the 
disk, $H$, we expect these to produce larger variations at 
smaller disk radii.

Recent calculations indicate that the inner regions of FUor disks 
may be much different than predicted by the simple disk model.
The steady-state temperature distribution assumes a physically thin
disk, $H \ll R$ (\cite{lyn74}).  FUor disks are 
probably much thicker (\cite{lin85}; \cite{cla89}, 1990).
Steady models that include a self-consistent treatment of the boundary 
layer between the inner disk and the stellar photosphere predict 
large scale heights, $H/R \sim$ 0.1--0.3 at $R \sim 1-2~R_{\star}$
(\cite{pop93}, 1996).  Time-dependent models further indicate 
that $H/R$ can vary in a complicated way close to the central 
star (\cite{tur97}; \cite{kle99}).  Both types of model predict 
that the disk temperature peaks just outside the stellar 
photosphere at $R \approx$ 1.1--1.2 $R_{\star}$.  The decline 
in disk temperature at smaller radii can be as large as 25\%--50\%.
Applied to FU Ori, these models predict that the disk temperature 
close to the central star is $\sim$ 5000--6000 K, comparable to the
temperature derived for the flickering source, if the peak 
temperature at 1.1--1.2 $R_{\star}$ 
is $\sim$ 7000 K.  

We propose that the flickering source in FU Ori lies between the 
stellar photosphere and the peak temperature in the disk at 1.1--1.2
$R_{\star}$.  In the models described above, this region produces 
$\sim$ 5\% of the total optical light.  Observable variations in 
the total light from the disk -- as we have reported here -- thus
imply significant changes in the physical structure of the inner 
disk.  Our data for FU Ori require 50\% variations in the light 
output of the inner disk.  

Large changes in the physical structure of the disk can be avoided
if the spatially thick portion of the disk occults the inner disk.
If we view the disk at an inclination $i_{crit}$ = tan$^{-1} H/R$,
small variations in $H/R$ can produce small changes in brightness
and color.  Rapid variations similar to those observed are possible
if $H/R$ varies on the dynamical timescale and if the occulted 
portion of the disk radiates as a G0 I star.  The required 
change in $H/R$, $\sim$ 10\%, is small compared to the 50\%
change in light output needed above.  The required geometry is,
however, very special and yields no variation if the real
viewing angle is much less than $i_{crit}$.  Observations of
V1057 Cyg and V1515 Cyg will test this idea, because these 
systems probably have smaller $i$ than FU Ori (\cite{har96}).

\section{DISCUSSION AND SUMMARY}

Our results provide the first evidence for rapid photometric variations, 
flickering, in a FUor.  The amplitude of the variation is small,
$\sim$ 0.035, and just detectable with photoelectric data covering
a long time interval.  Observations with smaller photometric 
uncertainties are needed to verify the detection and to place 
better limits on the color variations.  Differential photometry
using a CCD on a small telescope can achieve the required precision,
but the field of FU Ori has few bright comparison stars within
15--20 arcmin.  The richness of the field may compensate and
allow the high quality photometry needed to improve our results.

Previous attempts to find similar short-term variations in a 
pre--main-sequence star have met with mixed success.  Smith 
\etal (1996) placed upper limits of 0.01 mag on short-term fluctuations
in four classical T Tauri stars.  Gullbring \etal (1996) detected 
flare-like activity in the classical T Tauri star BP Tau, with 
amplitudes and timescales comparable to that observed in FU Ori; 
Hessman \& Guenther (1997) noted similar behavior in three classical
T Tauri stars, DG Tau, DR Tau, and DI Cep.  These studies all interpreted
the variations with a magnetospheric disk model, where jitter
in the magnetically channeled flow from the inner disk to the stellar
photosphere produces small amplitude `flares'.  We prefer to associate
the variation in FU Ori with flickering of the inner accretion disk.
The accretion rate in FU Ori, $\sim 10^{-4} \msunyr$, is a factor
of $\sim$ 1000 larger than accretion rates derived for T Tauri stars,
which makes it difficult to truncate the disk with the modest magnetic
fields, $\lesssim$ 1--2 kG, detected in pre-main sequence stars.
Future observations can test the magnetic alternative by placing 
better limits on any periodic component of the photometric variation 
and by measuring the magnetic field strength. 

However these observational issues are resolved, it is clear that
high precision photometry can probe the physical conditions of the
inner accretion disk of a pre--main-sequence star.  The amplitudes
and timescales of these variations already provide some challenge
to theory.  The amplitude of the FU Ori variation implies large
fluctuations in the physical structure of the disk on short 
timescales.  Flares and other short-term variations in T Tauri stars 
suggest smaller, but still significant, changes in disk structure
close to the central star.  Recent hydrodynamical calculations 
show that the disk structure can change significantly on longer 
timescales, but theoretical models do not yet address rapid 
fluctuations in the disk similar to those observed (\cite{kle99}).
Future calculations that consider this behavior should lead to 
a better understanding of mass flow in the inner disk in FUors
and other types of accreting systems.

\vskip 6ex

We thank F. Hessman for a careful and thoughtful review which 
improved our presentation of the data and our discussion of
possible models.
S.K. thanks N. Kylafis and C. Lada for the hospitality of the
NATO ASI, {\it The Origins of Stars and Planetary Systems.} 
Gentle Mediterranean waves rolling onto the beaches of Crete 
inspired portions of this study.

\vfill
\eject

\hskip -20ex
\epsffile{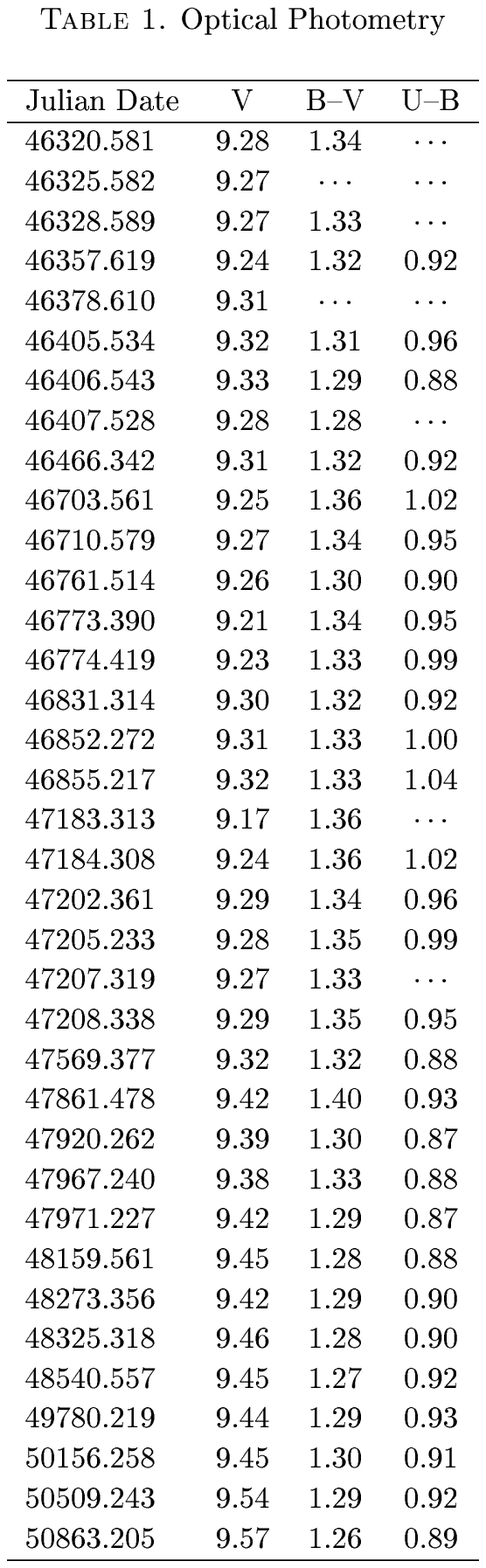}

\hskip -20ex
\epsffile{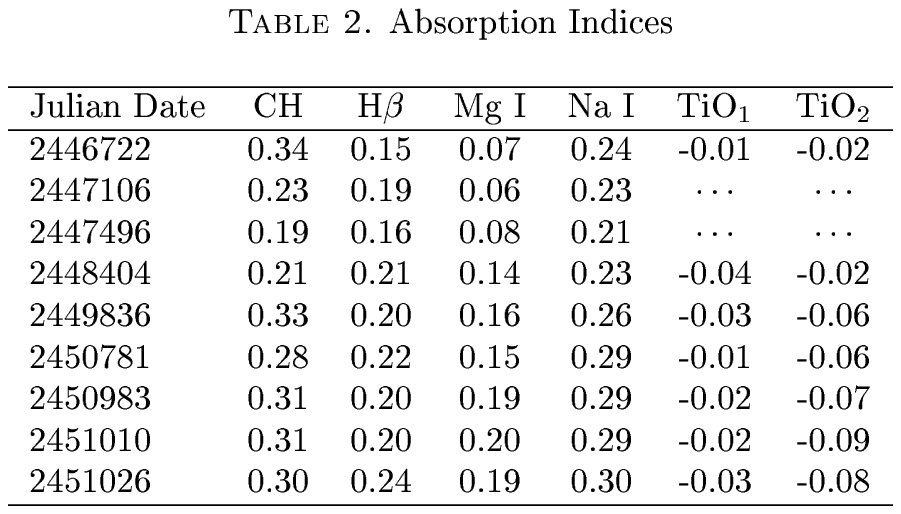}

\hskip -15ex
\epsffile{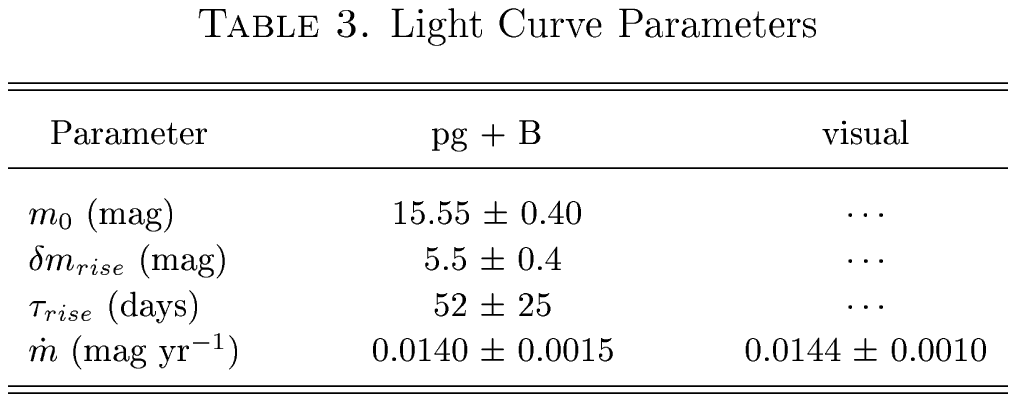}

\hskip 1ex
\epsffile{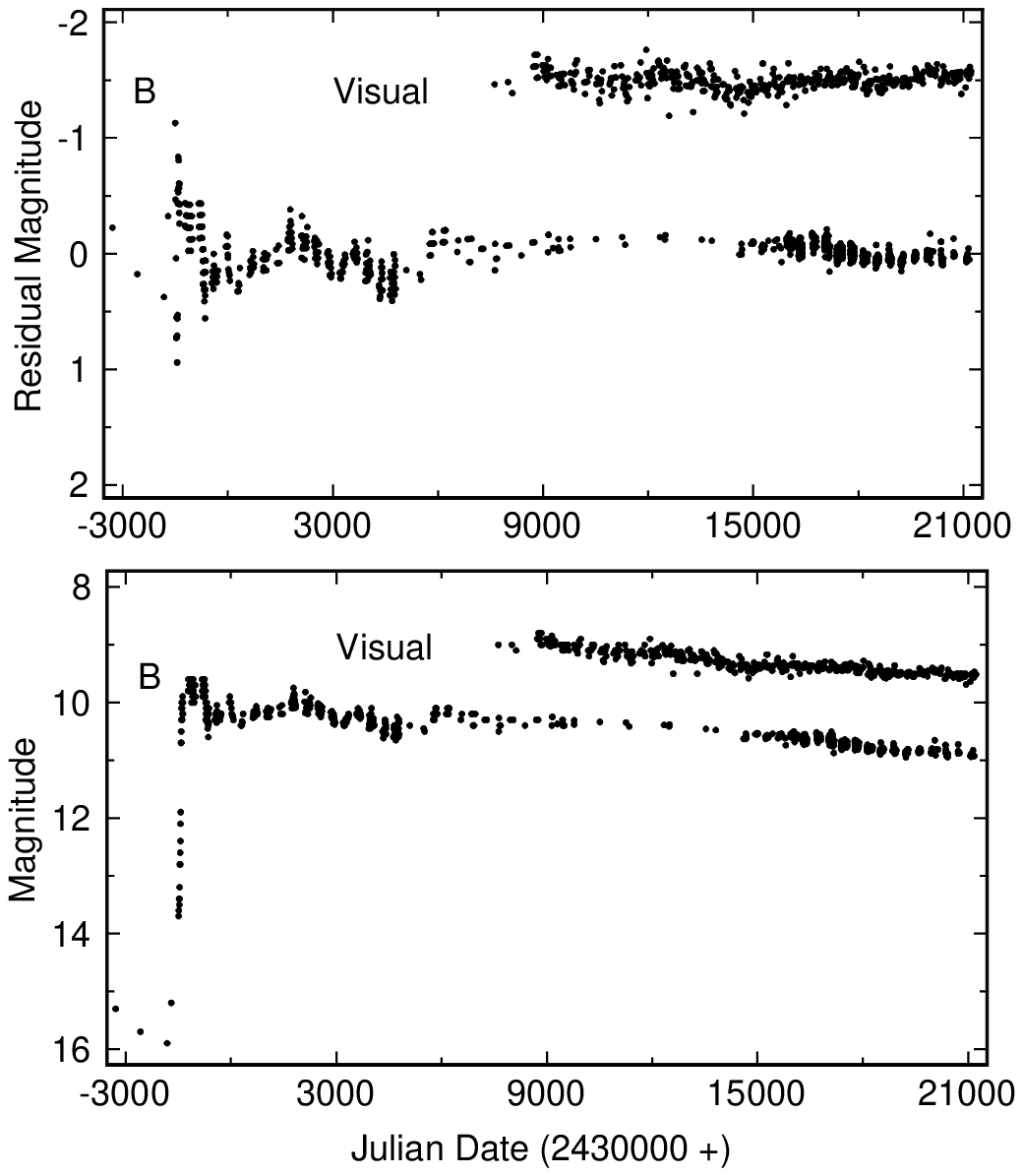}
\figcaption[Kenyon.fig1.ps]
{Optical (lower panel) and residual (upper panel) light curves 
for FU Ori.  The blue (B) light curve is comprised of 1042 
measurements derived from photographic plates and photoelectric 
photometry (Ibragimov 1997 and references therein).  The visual 
light curve is ten day means of 7300 visual observations from the 
AAVSO international database. The residuals plot the difference 
between the B and visual light curves in the lower panel 
and the model light curves described in the text. Both residual 
light curves show large, short timescale fluctuations superimposed 
on a wave-like variation with a timescale of 3000--3500 days.}

\hskip -7ex
\epsfxsize=8.0in
\epsffile{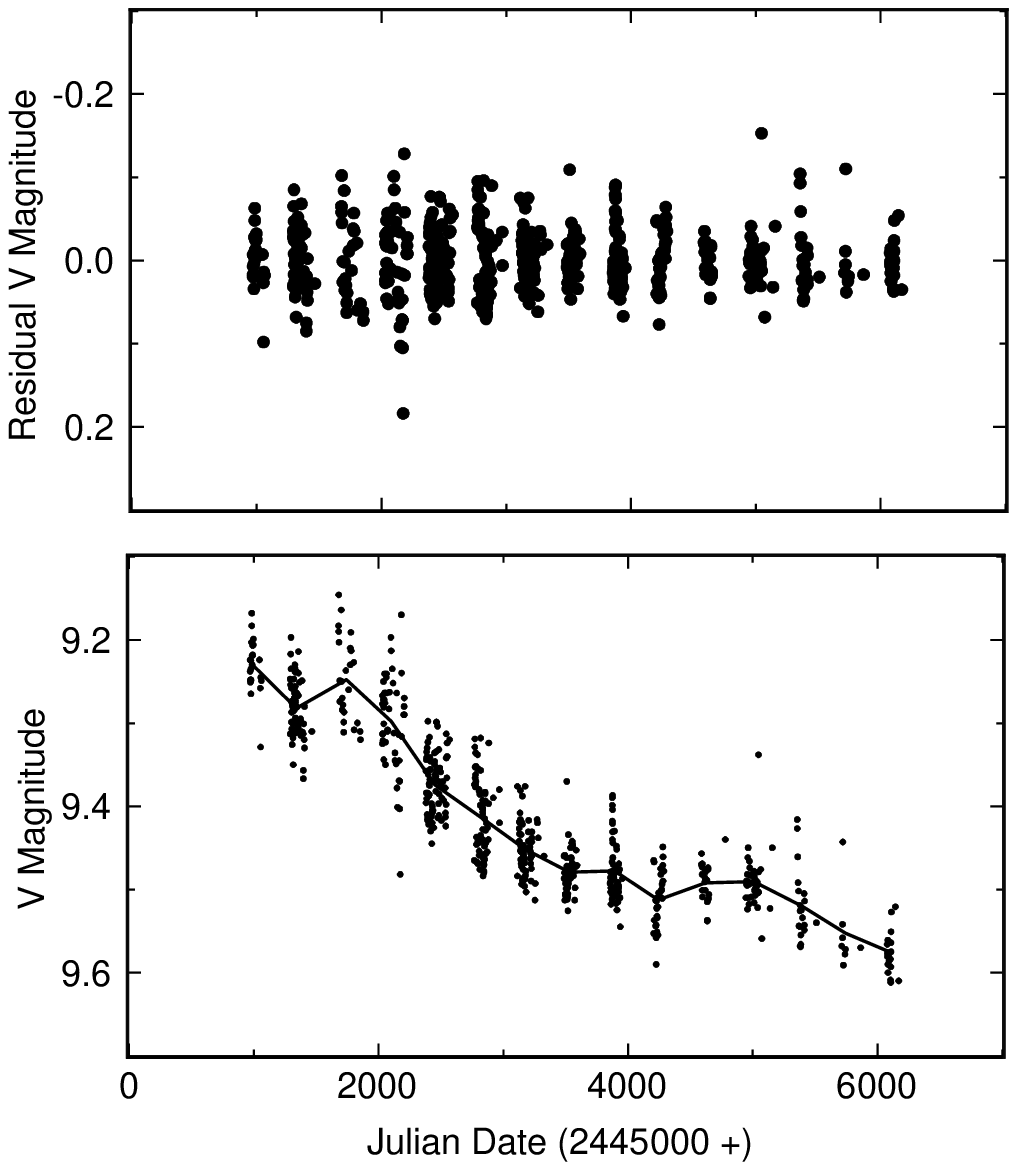}
\figcaption[Kenyon.fig2.ps]
{Photoelectric V-band data for FU Ori.
(a) bottom panel: filled circles indicate individual data points;
the solid line shows seasonal means. (b) top panel: variation of
individual data points about the seasonal mean light curve.}

\hskip -12ex
\epsfxsize=7.5in
\epsffile{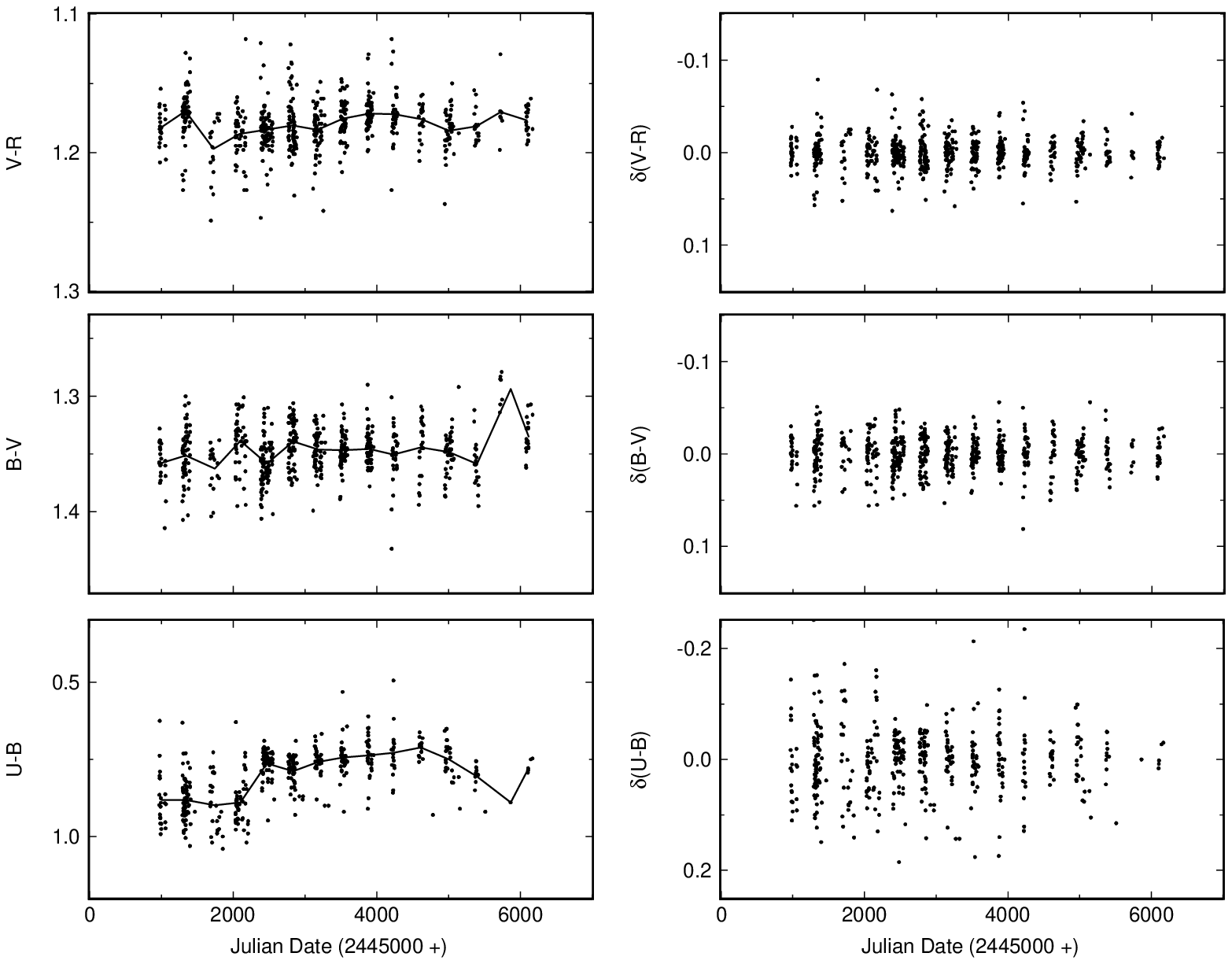}
\figcaption[Kenyon.fig3.ps] 
{Color variations in FU Ori. In each panel on the left, filled 
circles indicate individual data points; the solid line shows 
seasonal means.  Filled circles in the right hand panels 
indicate the variation of each color about the seasonal 
mean color curves in the left hand panels.}

\hskip -5ex 
\epsfxsize=7.0in
\epsffile{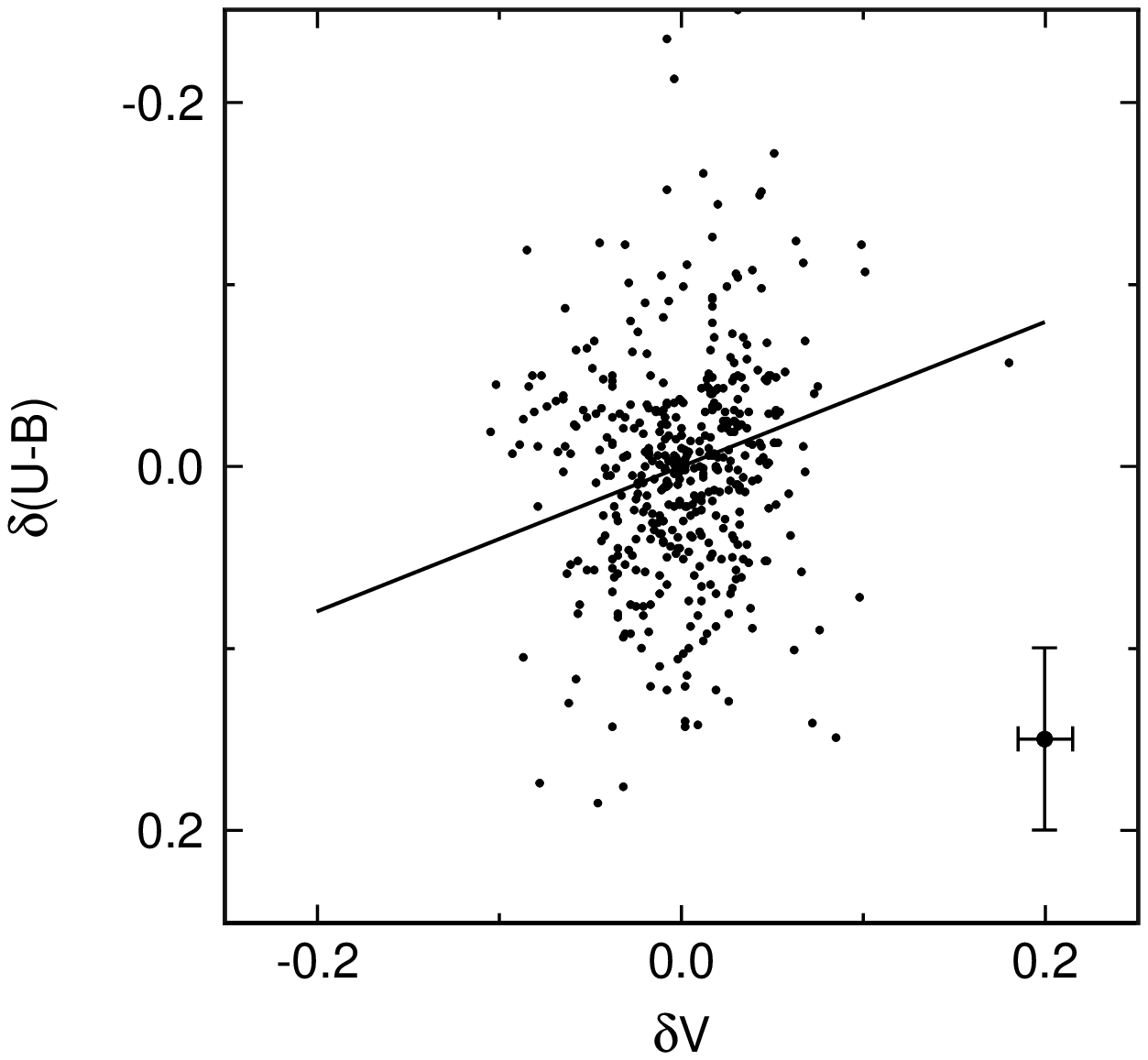}
\figcaption[Kenyon.fig4.ps]
{Correlation of U--B variations with V-band variations.
The straight line is a least squares fit to the plotted points.
The error bar in the lower right corner indicates the
uncertainty of a single observation.}

\hskip -10ex 
\epsfxsize=8.5in
\epsffile{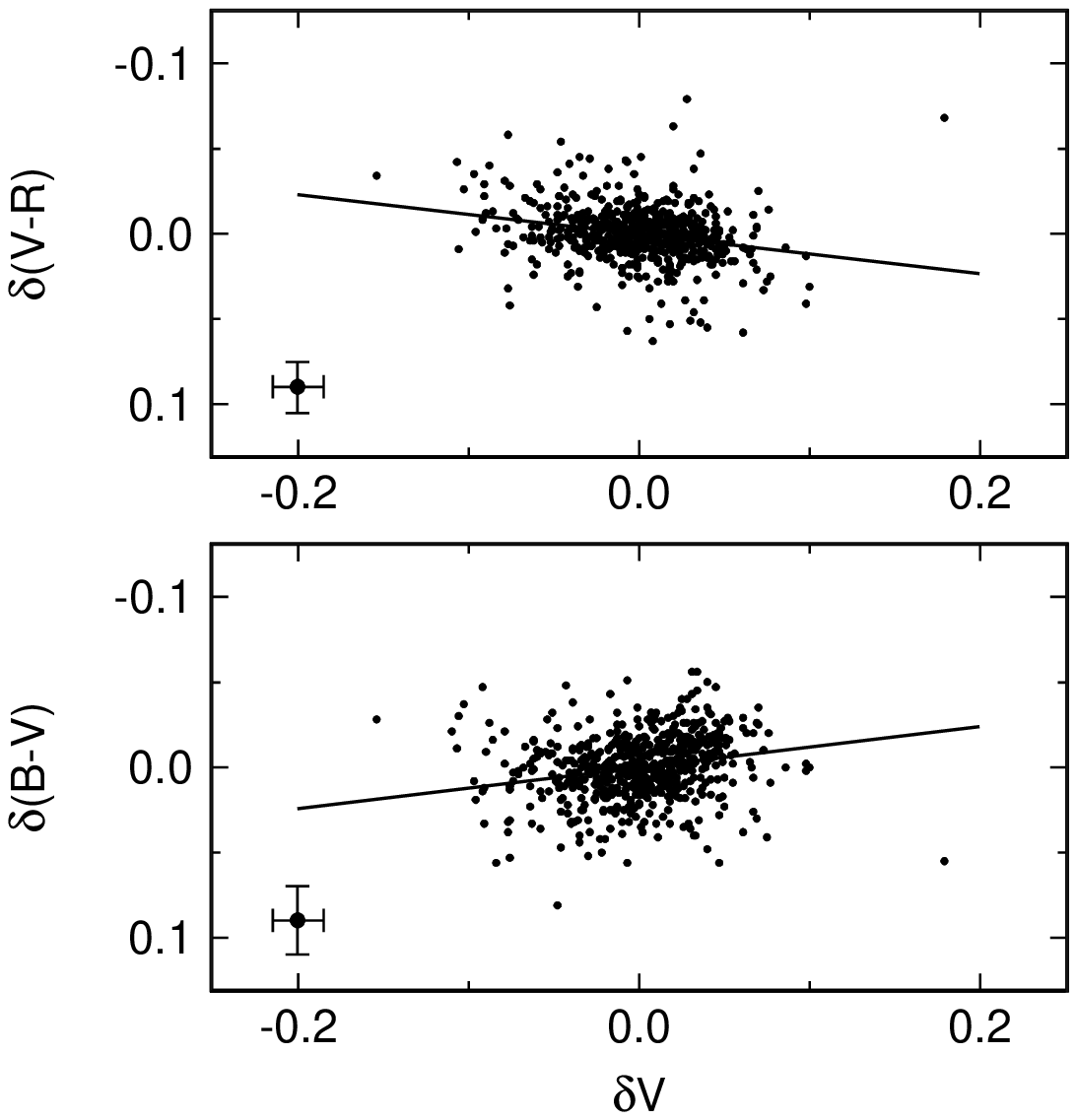}
\figcaption[Kenyon.fig5.ps]
{Correlation of color variations with V-band variations.
The lower panel plots the variation of B--V with V;
the upper panel plots the variation of V--R with V.
In each panel, the straight line is a least squares fit 
to the plotted points. The error bars indicate the
uncertainty of a single observation.}

\hskip -12ex
\epsfxsize=8.0in
\epsffile{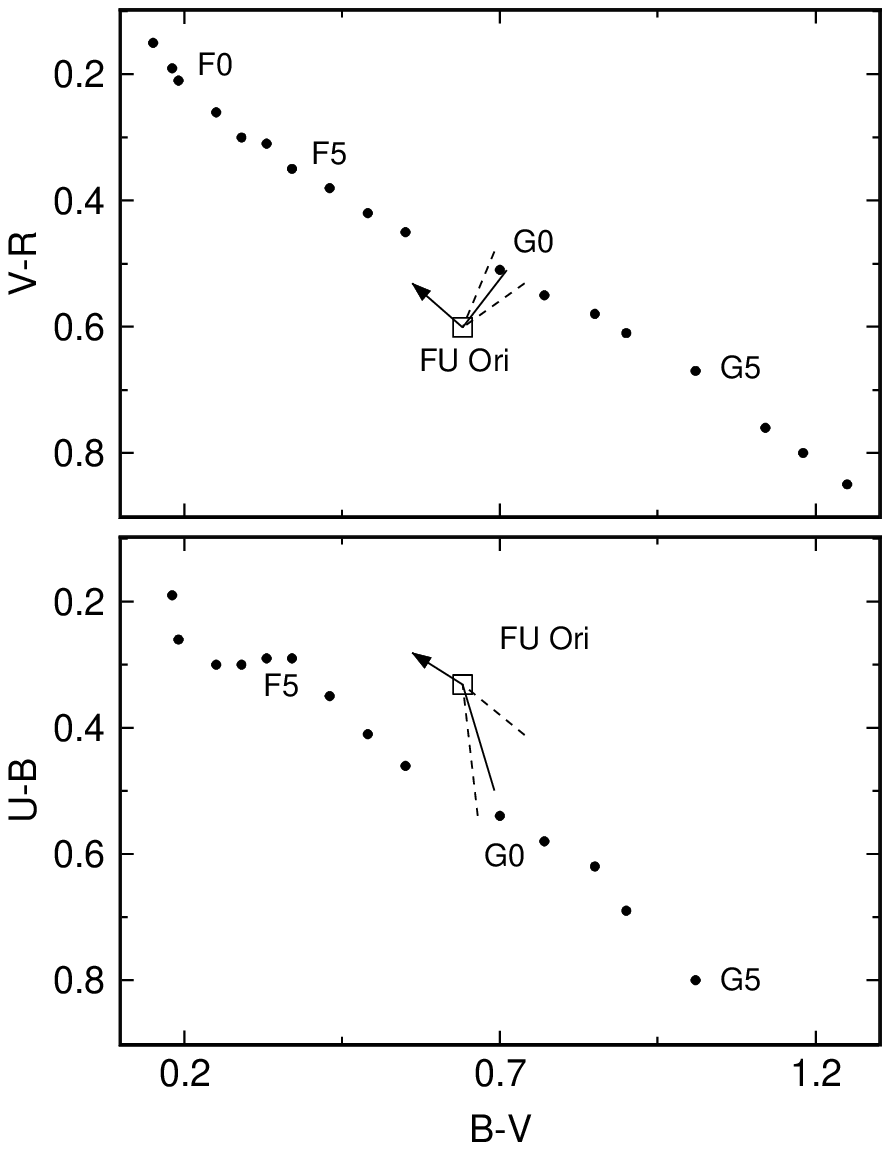}
\figcaption[Kenyon.fig6.ps]
{Flux ratio diagrams for FU Ori.  
In each panel, 
a box shows the reddening-corrected position of FU Ori,
filled circles show positions of F--G supergiants with 
indicated spectral types, the solid line points to the 
best stellar match to the colors of the variable source, and 
the dashed lines show $\pm 2 \sigma$ uncertainties in the
best stellar match.  A G0 supergiant provides the best match
to the observed color variations with an uncertainty of
roughly 1 spectral subclass.}


\clearpage

\begin{thebibliography}{}

\bibitem[Bell \etal 1995]{bel95} Bell, K. R., Lin, D. N. C., Hartmann, L., \& 
Kenyon, S. J. 1995, ApJ, 444, 376

\bibitem[Bruch 1992]{bru92} Bruch, A. 1992, A\&A, 266, 237

\bibitem[Bruch 1994]{bru94} Bruch, A. 1994, in
{\it Flares and Flashes}, IAU Colloquium No. 151, 
edited by J. Greiner, H. W. Duerbeck, \& R. E. Gershberg,
Berlin, Springer, p. 288

\bibitem[Bruch \& Duschl 1993]{bru93} Bruch, A., \& Duschl, W. J. 1993,
A\&A, 275, 219

\bibitem[Calvet \& Gullbring 1999]{cal99} Calvet, N. \& Gullbring, E. 1999,
ApJ, in press

\bibitem[Calvet \etal 1993]{cal93} Calvet, N., Hartmann, L., \& Kenyon, S. J. 
1993, ApJ, 402, 623

\bibitem[Carr \etal 1987]{car87} Carr, J. S., Harvey, P. M., \& Lester, D. F. 
1987, ApJL, 321, L71

\bibitem[Clarke \etal 1989]{cla89} Clarke, C. J., Lin, D. N. C., 
\& Papaloizou, J. C. B. 1989, MNRAS, 236, 495

\bibitem[Clarke \etal 1990]{cla90} Clarke, C. J., Lin, D. N. C., 
\& Pringle, J. E. 1990, MNRAS, 242, 439

\bibitem[Errico \etal 1997]{err97} Errico, L., Lamzin, S. A., Teodorani, M.,
Vittone, A. A., Giovanelli, F., \& Rossi, C. 1997, PAZh, 23, 687

\bibitem[Evans \etal 1994]{eva94}
Evans, II, N. J., Balkum, S., Levreault, R. M., Hartmann, L., 
\& Kenyon, S. J. 1994, ApJ, 424, 793

\bibitem[Fabricant \etal 1998]{fab98}  Fabricant, D. G., Cheimets, P., 
Caldwell, N., \& Geary, J. 1998, PASP, 110, 79

\bibitem[Feast \& Catchpole 1997]{fea97} Feast, M. W., \& Catchpole, R. M.
1997, MNRAS, 286, L1

\bibitem[Gautschy \& Saio 1995]{gau95} Gautschy, A., \& Saio, H. 1995,
ARA\&A, 33, 75

\bibitem[Goodrich 1987]{goo87} Goodrich, R. 1987, PASP, 99, 116

\bibitem[Gullbring \etal 1996]{gul96} Gullbring, E., Barwig, H., 
Chen, P. S., Gahm, G. F., \& Bao, M. X. 1996, A\&A, 307, 791

\bibitem[Hartmann \& Calvet 1995]{har95} Hartmann, L., \& Calvet, N. 
1995, AJ, 109, 1846

\bibitem[Hartmann \& Kenyon 1985]{har85} Hartmann, L., \& Kenyon, S.J. 
1985, ApJ, 299, 462

\bibitem[Hartmann \& Kenyon 1996]{har96} Hartmann, L., \& Kenyon, S. J. 
1996, ARA\&A, 34, 205

\bibitem[Hayes \& Latham 1975]{hay75} Hayes, D., and Latham, D. 1975, ApJ, 197, 593

\bibitem[Herbig 1966]{her66} Herbig, G. H. 1966, Vistas in Astr, 8, 109

\bibitem[Herbig 1977]{her77} Herbig, G. H. 1977, ApJ, 217, 693

\bibitem[Herbig \& Petrov 1992]{her92} Herbig, G. H., \& Petrov, P. P. 
1992, ApJ, 392, 209

\bibitem[Hessman \& Guenther 1997]{hes97} Hessman, F. V., \& 
Guenther, E. W. 1997, A\&A, 321, 497

\bibitem[Hindsley \& Bell 1989]{hin89} Hindsley, R. B. \& Bell, R. A. 
1989, ApJ, 341, 1004

\bibitem[Ibragimov 1997]{ibr97} Ibragimov, M. A. 1997, Pis'ma Astr. Zh., 
23, 125

\bibitem[Johnson 1966]{joh66} Johnson, H. L. 1966, ARA\&A, 3, 193

\bibitem[Johns-Krull \etal 1999]{joh99} Johns-Krull, C. M., Valenti, J. A.,
\& Koresko, C. 1999, ApJ, 516, 900

\bibitem[Kearns \etal 1997]{kea97} Kearns, K., Eaton, N. L., Herbst, W.,
\& Mazzurco, C. J. 1997, AJ, 114, 1098

\bibitem[Kenyon 1999]{ken99} Kenyon, S. J. 1999, in {\it The Origin of
Stars and Planetary Systems,} edited by C. J. Lada \& N. Kylafis, 
Dordrecht, Kluwer, in press

\bibitem[Kenyon \& Hartmann 1991]{ken91} Kenyon, S. J., \& Hartmann, L. 
1991, ApJ, 383, 664

\bibitem[Kenyon \etal 1988]{ken88} Kenyon, S. J., Hartmann, L., \& 
Hewett, R. 1988, ApJ, 325, 231

\bibitem[Kenyon \etal 1996]{ken96} Kenyon, S. J., Yi, I., \& Hartmann, L.
1996, ApJ, 462, 439

\bibitem[Kley \& Lin 1999]{kle99} Kley, W., \& Lin, D. N. C. 1999, 
ApJ, in press

\bibitem[Kolotilov \& Petrov 1985]{kol85} Kolotilov, E. A., 
\& Petrov, P. P. 1985, Pis'ma Astr. Zh., 11, 846

\bibitem[Kurtz \& Marang 1995]{kur95} Kurtz, D. W., \& 
Marang, F. 1995, MNRAS, 276, 191

\bibitem[Lamzin 1998]{lam98} Lamzin, S. A. 1998, Astr. Rept., 42, 322

\bibitem[Lin \& Papaloizou 1985]{lin85}
Lin, D. N. C., \& Papaloizou, J. 1985, in
{\it Protostars and Planets II,}
ed. D. C. Black and M. S. Matthews,
Tucson, University of Arizona Press, p. 981

\bibitem[Lynden-Bell \& Pringle 1974]{lyn74} Lynden-Bell, D., \& 
Pringle, J. E. 1974, MNRAS, 168, 603

\bibitem[Marconi \& Palla 1998]{mar98} Marconi, M., \& Palla, F. 1998,
ApJ, 507, L141

\bibitem[Mathis 1990]{mat90} Mathis, J. S. 1990, ARA\&A, 28, 37

\bibitem[Mould \etal 1978]{mou78} Mould, J. R., Hall, D. N. B., 
Ridgway, S. T., Hintzen, P., \& Aaronson, M. 1978, ApJL, 222, L123

\bibitem[Natta \etal 1997]{nat97} Natta, A., Grinin, V. P., Mannings, V.,
\& Ungerechts, H. 1997, ApJ, 491, 885

\bibitem[O'Connell 1973]{oco73} O'Connell, R.W. 1973, AJ, 78, 1074

\bibitem[Petrov \etal 1998]{pet98} Petrov, P., Duemmler, R., 
Ilyin, I., Tuominen, I. 1998, A\&A, 331, L53

\bibitem[Popham \etal 1993]{pop93} Popham, R., Narayan, R., 
Hartmann, L., \& Kenyon, S. J.  1993, ApJL, 415, L127

\bibitem[Popham \etal 1996]{pop96} Popham, R., Narayan, R., 
Kenyon, S. J., \& Hartmann, L. 1996, ApJ, 473, 422

\bibitem[Press \etal 1992]{pre92} Press, W. H., Flannery, B. P., 
Teukolsky, S. A., \& Vetterling, W. T. 1992, {\it Numerical Recipes, 
The Art of Scientific Computing,} Cambridge, Cambridge

\bibitem[Reipurth 1991]{rei91} Reipurth, B. 1991, in
{\it Physics of Star Formation and Early Stellar Evolution},
NATO Adv. Study Inst., edited by C. J. Lada and N. D. Kylafis, p. 497

\bibitem[Robinson 1976]{rob76} Robinson, E. L. 1976, ARA\&A, 14, 119

\bibitem[Rodriguez \etal 1990]{rod90} Rodriguez, L., Hartmann, L. W., 
\& Chavira, E. 1990, PASP, 102, 1413

\bibitem[Rodriguez \& Hartmann 1992]{rod92} Rodriguez, L., \& Hartmann, L. 
1992, Rev. Mex. A\&A, 24, 135

\bibitem[Rostopchina \etal 1997]{ros97} Rostopchina, A. N.,
Grinin, V. P., Okazaki, A., The, P. S., Kikuchi, S., Shakhovskoy, D. N.,
\& Minikhulov, N. Kh. 1997, A\&A, 327, 145

\bibitem[Smith \etal 1996]{smi96} Smith, K. W., Jones, D. H. P., \&
Clarke, C. J. 1996, MNRAS, 282, 167

\bibitem[Staude \& Neckel 1992]{sta92} Staude, H. J., \& Neckel, Th. 
1992, ApJ, 400, 556

\bibitem[Stocke \etal 1988]{sto88} Stocke, J. T., Hartigan, P. M., 
Strom, S. E., Strom, K. M., Anderson, E. R., Hartmann, L. W., \& 
Kenyon, S. J. 1988, ApJS, 68, 229

\bibitem[Turner \etal 1997]{tur97} Turner, N. J. J., Bodenheimer, P.,
\& Bell, K. R. 1997, ApJ, 480, 754

\bibitem[Warner 1995]{war95} Warner, B. {\it Cataclysmic Variable 
Stars,} Cambridge, Cambridge University Press

\bibitem[Weintraub \etal 1989]{wei89} Weintraub, D. A., Sandell, G.,
\& Duncan, W. D. 1989, ApJ, 340, 69

\bibitem[Weintraub \etal 1991]{wei91} Weintraub, D. A., Sandell, G.,
\& Duncan, W. D. 1991, ApJ, 382, 270

\end{thebibliography}
\end{document}